\documentclass[12pt]{article}
\usepackage{epsfig,amsfonts,amssymb}
\usepackage{amsmath}
\usepackage[margin=2.5cm]{geometry}
\usepackage{color}
\usepackage[colorlinks=true,urlcolor=blue,linkcolor=blue]{hyperref}
\usepackage{cite}
\usepackage{framed}
\usepackage{breqn}
\usepackage{flexisym}
\input epsf.sty
\usepackage{cite}

\def\ZZZ{{\hbox{ Z\kern-1.6mm Z}}}
\def\RRR{{\hbox{ R\kern-2.4mm R}}}
\def\CCC{{\hbox{ C\kern-2.0mm C}}}
\def\zzz{{\hbox{z\kern-1mm z}}}

\newcommand{\Tr}{\mathop{\mathrm{Tr}}}

\def\one{{\hbox{ 1\kern-.8mm l}}}
\def\zero{{\hbox{ 0\kern-1.5mm 0}}}

\begin{document}

\baselineskip 24pt

\begin{center}
{\Large \bf  Entanglement of Spatial Regions vs. Entanglement of Particles}

\end{center}

\vskip .6cm
\medskip

\vspace*{4.0ex}

\baselineskip=18pt

\centerline{\large \rm Swapnamay Mondal}

\vspace*{4.0ex}

\centerline{\large \it Harish-Chandra Research Institute}
\centerline{\large \it  Chhatnag Road, Jhunsi,
Allahabad 211019, India}

\vspace*{1.0ex}
\centerline{\small E-mail: swapno@hri.res.in}

\vspace*{5.0ex}

\centerline{\bf Abstract} \bigskip
Consider an arbitrary local quantum field theory with a gap or an arbitrary gapless free theory. We consider states in such a theory, that describe two entangled particles localized in disjoint regions of space. We show that in such a state, to leading order, R\'{e}nyi entropies of spatial regions, containing only one of the particles are same as their quantum mechanical counterparts, after subtraction of vacuum contribution. Subleading corrections depend on overlap of wave functions. These results suggest that Von Neumann entropy of a spatial region, after subtraction of vacuum contribution, can serve as a measure of entanglement of indistinguishable particles in pure states.

\vfill \eject

\baselineskip=18pt

\tableofcontents

\section{Introduction}  \label{s1}
What makes quantum mechanics mathematically nice is linearity. On the other hand what makes quantum mechanics physically  somewhat counterintuitive is also linearity. The best manifestation of this is entanglement, non triviality of which has been noted as early as 1935 \cite{EPR}. In last few decades it has also been realized that entanglement of  quantum particles can be put to practical use. This is at the  very heart of the subject of quantum information and communication. Recent years have witnessed widespread interest in another somewhat different aspect of entanglement, namely entanglement of spatial regions in quantum field theories. This has been found to be relevant in diverse areas of physics such as  holography \cite{Ryu:2006bv}, \cite{Ryu:2006ef}, entropic c theorem \cite{Casini:2004bw},\cite{casini2}, quantum phase transitions \cite{Vidal:2002rm},\cite{Kitaev:2005dm}, \cite{Wen} to name a few.

In this paper we ask a rather simple minded question, which is possibly of quite general interest. We consider a state in a general local quantum field theory that can be thought of as an entangled state of two localized particles. Then the entanglement (in the framework of quantum mechanics) of the emergent particles, is clearly a relevant notion in this context. On the other hand in the framework of field theory, the natural notion is entanglement of spatial regions (containing one of the localized particles). Hence one would expect some relation between these two notions. To our knowledge such a relation has not been investigate previously\footnote{R\'{e}nyi and Von Neumann entropies of  locally excited states have been studied in \cite{Caputa:2014vaa}, \cite{Nozaki:2014hna} for conformal field theories. But there one is mainly concerned about time evolution of entanglement. In \cite{Shiba:2014uia} mutual R\'{e}nyi information of locally excited entangled states was considered and it was noted that this admits a quantum mechanical interpretation. We show this for R\'{e}nyi entropies themselves and extract somewhat different physics than these papers.}. We address this question in this paper and show that such a relation exists for any local quantum field theory. Since field theory is a more fundamental description of elementary particles than quantum mechanics, this also provides a finer notion of entanglement of particles\footnote{In spirit, this is  similar to various physical quantities, e.g. magnetic moment of electron, which have a coarser description in quantum mechanics and a finer one in field theory. Although for entanglement the finery is an artefact of the state under consideration, rather than the theory.}.

Our results may have interesting bearing in black hole information paradox \cite{Hawking}, more precisely on Mathur's argument \cite{Mathur:2009hf} and AMPS argument \cite{Almheiri:2012rt}. This is because these arguments are based on quantum mechanical notion of entanglement, whereas the natural framework for the problem is quantum field theory in curved space times. To leading order, it would not matter, nevertheless it may be interesting to explore  small corrections due to this.

Our paper is organized as follows: in section \ref{s2}, we derive our main result (\ref{proofconjecture}), which is  relating R\'{e}nyi entropies of spatial regions to those of quantum particles. In \ref{s3} we discuss possible usage of our results for the problem of entanglement of indistinguishable particles.

\section{Entanglement: spatial regions vs particles}  \label{s2}
Consider a state of the following form
\begin{align}
 |\Psi\rangle 
 &=
 N
 \left( 
 a O_i \tilde{O}_j + b O_k \tilde{O}_l
 \right) |0\rangle\, .
 \label{psifield}
\end{align}
$N$ is a normalization constant and
$O_i$ denotes creation operator for a mode
corresponding to the spatial wave function
$\phi_1(x)$
 (which is concentrated around the point $x_1$) and 
internal index $i$, $\tilde{O}_j$ denotes creation operator for a mode
corresponding to the spatial wave function $\phi_2(x)$(which is concentrated around the point $x_2$) and internal index $j$. Similar statements hold for $O_k , \tilde{O}_l$.

For a wave function $\phi (x)$, or equivalently $\tilde{\phi}(k)$, the corresponding state is $\int \frac{d^D k}{\sqrt{2 E_k}} \tilde{\phi}(k) |k\rangle$, where $|k\rangle$ is the one particle state of momentum $k$.
For free theories (both gapped and gapless) we have explicit construction of such states in terms of creation operators.
For gapped interacting field theories, we still have unambiguous notion of one particle momentum eigenstates $|k\rangle$.	
In this case we can define ``creation operator" $a_k^\dagger$ by $|k\rangle =: a_k^\dagger |0\rangle$, with $|0\rangle$ being the interacting vacuum. Since one particle states are orthogonal to the bound states as well as continnum of multi-particle states, we have $a_k |\alpha \rangle = 0$ for any bound/ multi-particle state $|\alpha \rangle$. Thus $O,\tilde{O}$ are simply suitable linear superpositions of these $a_k^\dagger$-s. However the notion of one particle states is a bit ambiguous for gapless interacting theories. This is becasue being interacting one does not have an explicit creation operator to start with and being gapless it is not clear how to extract one particle states from continuous spectrum. Hence we would exclude such theories form our analysis.

We choose the following normalization\footnote{Note as $O$-s
contain only creation operators, they (anti)commute among
themselves.} for $O$-s
\begin{align}
\langle 0 | O_i^\dagger O_j |0\rangle = \delta_{ij}
= \langle 0 | \tilde{O}_i^\dagger \tilde{O}_j |0\rangle \, .
 \label{normalization}
\end{align}
Now if the wave functions $\phi_1(x)$ and $\phi_2(x)$ are 
entirely supported in disjoint regions $A_1$ and $A_2$,
such a state describes an entangled state of two particles,
distinguishable by their positions. This may then be 
described, within the framework of quantum mechanics of 
distinguishable particles, as 
\begin{align}
 |\psi_{spin} \rangle 
 &=
 \frac{1}{\sqrt{|a|^2 + |b|^2}}
 \left(
 a |i \rangle |j \rangle
 +
 b |k \rangle |l \rangle
 \right)\, ,
 \label{psiqm}
\end{align}
with the spatial wave functions $\phi_1(x)$ and $\phi_2(x)$ serving as particle 
labels.

A natural question to ask is that what is the connection between 
entanglement of \ref{psiqm} and entanglement of the region say $A_1$,
in the state \ref{psifield}? We answer this
question in this paper for arbitrary local quantum field theory with a gap and for arbitrary local gapless free theory.

	Bipartite entanglement of a pure state is quantified by the Von Nuemann entropy $S(\rho) = - \Tr{} \rho \log{} \rho$, where $\rho$ is the density matrix of any of the parties. However 
	for our purpose, R\'{e}nyi entropies prove more useful. $n^{th}$ R\'{e}nyi entropy of a density matrix $\rho$ is given by 
	\begin{align}
		S_n(\rho)
		&=
		\frac{1}{1-n} \log{} \Tr{} \rho^n \, ,
		\label{renyiDef}
	\end{align}
	where $\rho$ is the density matrix of any of the parties. A party would stand for a paricle (or a collection of particles) while discussing a quantum mechanical state and a region of space while discussing a field theory state.
	
	For computational reasons it is even more advantageous to look at the following quantity
	\begin{align}
	R_n\left(\rho\right)
	& :=
	e^{(1-n)S_n\left(\rho\right)}
	=
	\Tr{} \rho^n \, .
	\label{RnDef}
	\end{align}
First we note that we can express $R_n$ as
\begin{align}
R_n(\rho) 
& = 
\Tr{}_n
\left[
\rho^{\otimes n} E^{(n)}
\right]\, , 
\label{RnEn}
\end{align}
with
\begin{align}
E^{(n)}_{j_1\dots j_n;i_1 \dots i_n}
&= 
\prod_{k=1}^n \delta_{j_k,i_{k+1}}\, .
\label{En}
\end{align}	
$\Tr{}_n$ stands for trace over $\mathcal{H}_1^{\otimes n}$, where $\mathcal{H}_1$ is the Hilbert space of the subsystem under consideration. The indices $\{i_k,j_k\}$ take discreet values if we are discussing a quantum mechanical state and continuous values if we are discussing a field theoretic state. Note that \ref{RnEn} does not fix $E^{(n)}$  uniquely.
	
%

As defined above, $E^{(n)}$ is a linear operator 
on $\mathcal{H}_1^{\otimes n}$.
One can easily extend its action
on the full tensor product Hilbert space 
$\left(\mathcal{H}_1 \otimes \mathcal{H}_2\right)^{\otimes n}$
by defining
\begin{align*}
 E^{(n)}_{j_1, b_1\dots j_n,b_n;i_1,a_1 \dots i_n,a_n}
 &= 
 \prod_{k=1}^n \delta_{j_k,i_{k+1}}
 \delta_{a_k,b_k}\, ,
\end{align*}
where $i_k,j_k$ are the indices of $k$-th copy of 
$\mathcal{H}_1$ and 
$a_k,b_k$ are the indices of $k$-th copy of 
$\mathcal{H}_2$. 
In fact if $\rho_1 = \Tr_2 \rho_{12}$, then
\begin{align*}
 R_n (\rho_1) &= \Tr{}_n \left[ \rho_{12}^{\otimes n}
 E^{(n)}\right]\, ,
\end{align*}
where now trace is over
$\left(\mathcal{H}_1 \otimes \mathcal{H}_2\right)^{\otimes n}$.
From now on we will suppress the subscripts on trace and assume 
it is clear from the context.

For us $\rho_{12} = |\Psi\rangle \langle \Psi|$ (with $|\Psi\rangle$ given by \ref{psifield}) and subsystem that we are interested in, is a spatial region  $\Omega$. Thus
$\rho_1 = \rho_\Omega = \Tr_{\Omega_c} |\Psi\rangle \langle \Psi|$,  $\Omega_c$ being the complementary region of $\Omega$.
The operator $E^{(n)}_\Omega$
has been studied in great details in a recent paper
\cite{Shiba:2014uia} and we will heavily borrow their results.
If we denote local field variable at point $a$ as $q_a$,
then $E^{(n)}_\Omega$ reads
\begin{align}
 E_\Omega^{(n)} 
 &=
 \int \prod_{j=1}^n \prod_{a \in \Omega} dq^{(j)}_a
 \prod_{b \in \Omega} dq^{(j')}_b
 |  
 \{q^{1'}_b \dots q^{n'}_b\}
 \rangle
 \langle
 \{q^{1}_a \dots q^{n}_a\}
 |~
 \delta \left(q^{(1')}_b - q^{(2)}_a\right)
 \dots
 \delta \left(q^{(n')}_b - q^{(1)}_a\right)
 \times
 I_{\Omega_c}^n\, ,
 \label{eexpression}
\end{align}
where $I_{\Omega_c}^n$ is identity operator on $\mathcal{H}_{\Omega_c}^{\otimes n}$, $\mathcal{H}_{\Omega_c}$ being the Hilbert space associated with field variables living in the region $\Omega_c$. Perhaps a more comprehensible way to express $E_\Omega^{(n)}$ is the following
	\begin{align}
	\nonumber
		E_\Omega^{(n)}
		&=
		\int D\phi_\Omega^1 \dots D\phi_\Omega^n ~ D\phi_\Omega^{1'} \dots D\phi_\Omega^{n'}~ D\phi_{\Omega_c}^1 \dots D\phi_{\Omega_c}^n
		\delta (\phi_\Omega^{1'} - \phi_\Omega^2) \dots \delta (\phi_\Omega^{n'} - \phi_\Omega^1)
		\\
		\nonumber
		&
		| \phi_\Omega^{1'} \dots \phi_\Omega^{n'} \rangle 
		\langle \phi_\Omega^{1} \dots \phi_\Omega^{n} |
		\otimes
		| \phi_{\Omega_c}^{1} \dots \phi_{\Omega_c}^{n} \rangle 
		\langle \phi_{\Omega_c}^{1} \dots \phi_{\Omega_c}^{n} |
		\\
		&=
	\int D\phi_\Omega^1 \dots D\phi_\Omega^n ~ D\phi_{\Omega_c}^1 \dots D\phi_{\Omega_c}^n
	| \phi_\Omega^{2} \dots \phi_\Omega^{1} \rangle 
	\langle \phi_\Omega^{1} \dots \phi_\Omega^{n} |
	\otimes
		| \phi_{\Omega_c}^{1} \dots \phi_{\Omega_c}^{n} \rangle 
		\langle \phi_{\Omega_c}^{1} \dots \phi_{\Omega_c}^{n} | \,.
		\label{compE}
	\end{align}
	Here $|\phi_\Omega\rangle$ denotes the state corresponding to a smooth field configuration over region $\Omega$ and similar statement holds for  $|\phi_{\Omega_c}\rangle$. These are analogs of position eigenstates in quantum mechanics.
$ E_\Omega^{(n)} $ has many nice properties, as explored in \cite{Shiba:2014uia}.
The one particularly useful for us is
	\begin{align}
			\langle \psi_1 \dots \psi_n | E_\Omega^{(n)} | \chi_1 \dots \chi_n \rangle
			&=
			\langle \chi_{2} \dots \chi_1 | E_{\Omega_c}^{(n)} | \psi_1 \dots \psi_n \rangle^*\, .
			\label{pproperty}
	\end{align}
A derivation of this property is given in \ref{a1}.

Armed with these, we set out to compute $R_n(\rho_\Omega)$.
We consider a bosonic theory to start with. The main result of this section, namely  \ref{proofconjecture} would hold both for bosonic and fermionic fields. The corrections to this (disussed in \ref{a2}) would differ though. First we note
\begin{align*}
 R_n(\rho_\Omega)
 &=
 \Tr_{} \left( 
 |\Psi^{\otimes n}\rangle \langle \Psi^{\otimes n} |
 E^{(n)}_\Omega \right)
 =
 \langle \Psi^{\otimes n} | E_\Omega^{(n)} | \Psi^{\otimes n} \rangle \, .
\end{align*}
This contains $4^n$ terms, each of the following form
\begin{align*}
 &
  \langle 0 |
\left( \tilde{O}^{(n)}_{j_n} \right)^\dagger 
\left( O^{(n)}_{i_n}\right)^\dagger
  \dots
\left( \tilde{O}^{(n)}_{j_n}\right)^\dagger 
\left( O^{(n)}_{i_n}\right)^\dagger
  E_\Omega^{(n)}
  O^{(1)}_{p_1} \tilde{O}^{(1)}_{q_1}
  \dots
    O^{(1)}_{p_1} \tilde{O}^{(1)}_{q_1}
  | 0 \rangle  \\
 &=
  \langle 0|
    \left\{ \left( \tilde{O}^{(n)}_{j_n}\right)^\dagger
    \dots
    \left( \tilde{O}^{(1)}_{j_1}\right)^\dagger \right\}
   \left\{ \left( O^{(n)}_{i_n}\right)^\dagger
  \dots
  \left( O^{(1)}_{i_1}\right)^\dagger \right\}
  E_\Omega^{(n)}
   \left\{ O^{(1)}_{p_1}
  \dots
  O^{(n)}_{p_n} \right\}
  \left\{ \tilde{O}^{(1)}_{q_1}
  \dots
  \tilde{O}^{(n)}_{q_n} \right\}
  | 0 \rangle\, .
 \end{align*}
Here we have ignored the normalization factor $N^{2n}$.
To shorten the expressions, we introduce some notations
\begin{align}
\nonumber
  \left( O^{(n)}_{i_n}\right)^\dagger
  \dots
  \left( O^{(1)}_{i_1}\right)^\dagger
  &= :
  O_i^\dagger \\
  \nonumber
  \left( \tilde{O}^{(n)}_{j_n}\right)^\dagger
  \dots
  \left( \tilde{O}^{(1)}_{j_1}\right)^\dagger 
  &=:
  \tilde{O}_j^\dagger \\
  \nonumber
  O^{(1)}_{p_1}
  \dots
  O^{(n)}_{p_n}
  &=:
  O_p \\
  \tilde{O}^{(1)}_{q_1}
  \dots
  \tilde{O}^{(n)}_{q_n}
  &=:
  O_q\, .
  \label{notation}
\end{align}
So we have
\begin{align}
 \langle 0 |
 \tilde{O}_j^\dagger O_i^\dagger 
 E_\Omega^{(n)}
 O_p \tilde{O}_q |0\rangle\, .
 \label{genericterm}
\end{align}
Remember, in present notation all the $O,O^\dagger$-s are
clusters of $n$ creation/annihilation operators.
After some more steps, which we describe in appendix \ref{a2}, to set the clutter aside, this can be written as
	\begin{align}
			 	\langle 0| E_\Omega^{(n)} |0\rangle \times
			 	\delta_{p_2,i_1} \dots \delta_{p_1,i_n} 
			 	\times \delta_{j_1 q_1} \dots \delta_{j_n q_n}		
	+
	\text{other pieces}\, .
	\label{correctionterms}
	\end{align}
``Other pieces" are discussed in detail in appendix \ref{a2}. For now we just mention the relevant property of these pieces. They are small when the region $\Omega$ is chosen such that $\phi_1(x)$ is mostly supported inside $\Omega$ and $\phi_2(x)$ is supported mostly outside $\Omega$. Thus we concentrate on the leading piece.
When this piece is added for all the $4^n$ terms, one ends up with
\begin{align}
 N^{2n} \langle 0 | E_\Omega^{(n)} | 0 \rangle
 \left( |a|^{2n} + |b|^{2n} \right)
 &= 
 \langle 0 | E_\Omega^{(n)} | 0 \rangle
 \left( N \sqrt{|a^2| + |b|^2} \right)^{2n}
 R_n(\rho_1^{QM})\, ,
 \label{rho1qm}
\end{align}
where $\rho_1^{QM}$ is the density matrix of the first particle in the state \ref{psiqm}.
If $O$ and $\tilde{O}$ are supported in disjoint regions, which is the case considered,
$N \sim (\sqrt{|a|^2 + |b|^2})^{-1}$.
Thus
\begin{align}
\nonumber
 R_n\left(\rho_\Omega (\Psi)\right)
	&\sim
	R_n\left(\rho_\Omega (0)\right)
	\times R_n \left( \rho_1^{QM} (\psi) \right)
	\\
	\text{or,~~}
 S_n\left(\rho_\Omega (\Psi)\right)
 -
 S_n\left(\rho_\Omega (0)\right)
 &\sim
 S_n \left( \rho_1^{QM} (\psi) \right)\, ,
 \label{proofconjecture}
\end{align}
to leading order. 
 ``Other pieces" mentioned in \ref{correctionterms} and deviation of 
$N$ from $(|a^2| + |b|^2)^{-1/2}$ constitute corrections
to this.
We mention one interestig property of one of the correction terms (see \ref{a2} for detail).  Among ``other pieces", second piece of \ref{even1} contains a factor of
	$\prod_k j_k p_k$. This will 
	give non-zero contribution if $i = l$ or $j=k$ or $i=j$ or $k=l$
	in \ref{psifield}. 
	
A little thought would convince the reader that if we consider a more general state than \ref{psifield}, i.e. a state of the form
\begin{align}
	|\Psi\rangle
	&=
	N \sum_\alpha a_\alpha O_{i_\alpha} \tilde{O}_{ j_\alpha} |0\rangle\ \, ,
\end{align}
\ref{proofconjecture} would still hold. In fact our results continue to hold when instead of a single particle each subsystem contains a bunch of particles localized in some region. Now $O,\tilde{O}$ would contain a bunch of creation operators, but that does not change any of the steps.

Another interesting case is of occupation number entanglement. In this case, one would consider states of the form 
\begin{align}
	|\Psi\rangle
	&=
	\sum_n a_n O^n \tilde{O}^n |0\rangle \, .
\end{align}
In occupation number eigenbasis (and assuming orthogonality of the modes occupied) the above state would read
\begin{align}
	|\psi\rangle
	&=
	\sum_n a_n |n,n\rangle \, .
\end{align} 
We can repeat the very same steps as in \ref{a2}. Corresponding to each Kr\"{o}necker delta in spin indices in \ref{a2}, now we would have a Kr\"{o}necker delta in occupation number. Hence for present purpose they play the role of spin indices and \ref{proofconjecture} alongwith correction terms also holds for occupation number entanglement.
\section{Discussion} \label{s3}
These results prove useful for a problem in quantum mechanics, namely the problem of entanglement of indistinguishable particles\footnote{Many attempts have been made in recent years to generalize the notion of entanglement for indistinguishable particles	\cite{zarnardi},\cite{zarnardiandwang},	\cite{eckertetal}, \cite{yushi},\cite{korbiczandlewenstein}, \cite{Balachandran:2013hga}. Look at \cite{review1},\cite{review2} for recent review and references there in. These approaches mostly explore two directions, namely	tensor product structure of the Hilbert space and occupation number entanglement between various modes (which is indeed well suited for many systems considered in laboratory).}. If wave functions of various particles are nearly orthogonal they can be assigned different values of some observable and the particles can effectively be labelled by the value of that observable. Afterwards one can discuss entanglement of those particles  treating them as distinguishable particles. E.g. electrons stuck to various lattice sites can be labelled by the site it sits on (i.e. position) and they can be entangled through spin.

When wave functions start overlaping such labelling becomes ambiguous and the inherently indistinguishable nature of fundamental particles become important. The primary difficulty in discussing entanglement in such situation is identifying distinguishable subsystems, for particles are no more distinguishable. It is useful to take a field theoretic view of the situation, for then spatial regions serve as natural subsystems. When wavefunctions are localized around different positions, one can choose a region $\Omega$ containing only a single particle. The subsystem $\Omega$ then corresponds to that particle. This can be made quantitative if one finds a field theoretic quantity $\tilde{S}$, which is nearly equal to the Von Neumann entropy of the quantum mechanical density matrix of that particle. Even if the wave functions overlap, $\tilde{S}$ remains well defined. Thus provided $\tilde{S}$ is finite, it would be a natural candidate for entanglement of indistinguishable particles. The subsystem $\Omega$ can not clearly be associated with any particle though.

\ref{proofconjecture} suggests $S(\rho_\Omega (\Psi)) - S(\rho_\Omega (0))$ is the natural candidate for $\tilde{S}$, where we have used notations of last section and $S(\rho)$ denotes the Von Neumann entropy of the density matrix $\rho$. This quantity has the correct limit as the wavefunction overlap approaches zero. Given that Von Neumann entropy quantifies entanglement in pure states \cite{bennett}, $S(\rho_\Omega (\Psi))$ indeed quantifies entanglement. Further subtraction of vacuum contribution renders it finite.

\ref{proofconjecture} also tells that $S(\rho_\Omega (\Psi)) - S(\rho_\Omega (0))$ would in general differ from the naive quantum mechanical answer $S$, the difference being given by the ``correction terms" described in \ref{a2} (for R\'{e}nyi entropies). It should be noted that these corrections are never exactly zero, since there are no localized particles in a local quantum field theory \cite{reeh-schlieder}. We interpret these corrections as field theoretic corrections to naive quantum mechanical notion of entanglement. Due to these corrections, one would in general require an infinite dimensional density matrix to account for the vaccum subtracted Von Neumann and R\'{e}nyi entropies. In the spectrum of this density matrix field theoretic corrections correspond to infnitely many small eigenvalues. In terms of the entanglement spectrum\footnote{Modular Hamiltonian $H$ for a given density matrix $\rho$ is defined as $\rho = e^{-H}$. Previously the spectrum of $H$, called ``entaglement spectrum", has been argued to carry footprints of topological order\cite{haldane}, in the context of fractional quantum Hall effect. The low energy states of the entanglement spectrum become gapless, as the system becomes topologically ordered.}, this means field theoretic effects are encoded in ``high energy states", i.e. quantum mechanics serves as a ``low energy effective theory" in the context of entanglement!

These correction terms also lead to interesting properties for the lattice analog of $\tilde{S}$, namely site entanglement \cite{zarnardi},\cite{zarnardiandwang}. E.g. a single electron with wavefunction supported at more than one lattice site looks like an entangled state\footnote{Another strange feature is that the entanglement (through spin) of two electrons at same site (distinguished by orbitals quantum number) is invisible to site entanglement. These points have previously been noted in \cite{zarnardi}, \cite{grittings-fisher}.  Also look at \cite{wiseman-vaccaro},\cite{dowling-doherty-wiseman}.}! However bizarre, such entanglement has actually been used in teleportation \cite{lee-kim} and therefore physical. Given this it is natural to wonder whether the continuum analog of this can have some practical use as well. E.g. due to the field theoretic corrections, entanglement of a ``single" electron can exceed $\log{} 2$! It would be interesting to explore possible implications of this for quantum information and communication\footnote{We thank Arunabha Saha for pointing out this possibility.}.

\noindent {\bf Acknowledgements:} 
I am indebted to my friend Pinaki Banerjee for collaboration in earlier part of this work and many subsequent discussions. My special thanks to Ashoke Sen for various illuminating discussions and valuable comments on the manuscript. I would also like to thank Nilay Kundu, Arunabha Saha and whole QIC group at HRI,
especially H.S. Dhar, M.N. Bera, Aditi Sen(De) and Avijit Misra for useful conversations and comments. Lastly I thank people of India for their generous support for research in theoretical physics. This work has been supported by DAE project 12-R\&D-HRI-5.02-0303.

\appendix
\section{Derivation of \ref{pproperty}} \label{a1}
To make the logic more tractable we derive \ref{pproperty} for $n=3$. 
	\begin{align*}
	E_\Omega^{(3)}
	&=
		\int D\phi_\Omega^{1'} D\phi_\Omega^{2'} D\phi_\Omega^{3'} D\phi_\Omega^{1} D\phi_\Omega^{2} D\phi_\Omega^{3}~
		\delta(\phi_\Omega^{1'} - \phi_\Omega^{2}) \delta(\phi_\Omega^{2'} - \phi_\Omega^{3}) \delta(\phi_\Omega^{3'} - \phi_\Omega^{1})
		|\phi_\Omega^{1'} \phi_\Omega^{2'} \phi_\Omega^{3'} \rangle
		\langle \phi_\Omega^{1} \phi_\Omega^{2} \phi_\Omega^{3} |
		\\
		&
		\times
\int D\phi_{\Omega_c}^{1} D\phi_{\Omega_c}^{2} D\phi_{\Omega_c}^{3}
|\phi_{\Omega_c}^{1} \phi_{\Omega_c}^{2} \phi_{\Omega_c}^{3} \rangle
\langle \phi_{\Omega_c}^{1} \phi_{\Omega_c}^{2} \phi_{\Omega_c}^{3} |
\\
&=
\int  D\phi_\Omega^{1} D\phi_\Omega^{2} D\phi_\Omega^{3}~
D\phi_{\Omega_c}^{1} D\phi_{\Omega_c}^{2} D\phi_{\Omega_c}^{3}~|\phi_\Omega^{2} \phi_\Omega^{3} \phi_\Omega^{1} \rangle
\langle \phi_\Omega^{1} \phi_\Omega^{2} \phi_\Omega^{3} |
\times
|\phi_{\Omega_c}^{1} \phi_{\Omega_c}^{2} \phi_{\Omega_c}^{3} \rangle
\langle \phi_{\Omega_c}^{1} \phi_{\Omega_c}^{2} \phi_{\Omega_c}^{3} | \, .
	\end{align*}
	Now we can express any state $|\psi\rangle$ as
	\begin{align*}
		|\psi\rangle
		&=
		\int D\phi_\Omega D\phi_{\Omega_c}
		\psi[\phi_\Omega , \phi_{\Omega_c}]
		|\phi_\Omega\rangle \otimes |\phi_{\Omega_c}\rangle \, .
	\end{align*} 
	Hence
	\begin{align*}
		\langle \psi_1 \psi_2 \psi_3 | E_\Omega^{(3)} |\chi_1 \chi_2 \chi_3 \rangle
		&=
		\int  D\phi_\Omega^{1} D\phi_\Omega^{2} D\phi_\Omega^{3}~
		D\phi_{\Omega_c}^{1} D\phi_{\Omega_c}^{2} D\phi_{\Omega_c}^{3}~
		\\
		&
		\psi_1^*[\phi_\Omega^2,\phi_{\Omega_c}^1]~
		\psi_2^*[\phi_\Omega^3,\phi_{\Omega_c}^2]~
		\psi_3^*[\phi_\Omega^1,\phi_{\Omega_c}^3]~
		\chi_1 [\phi_\Omega^1,\phi_{\Omega_c}^1]~
		\chi_2 [\phi_\Omega^2,\phi_{\Omega_c}^2]~
		\chi_3 [\phi_\Omega^3,\phi_{\Omega_c}^3]
		\\
		&=
		\Big[ 
				\int  D\phi_\Omega^{1} D\phi_\Omega^{2} D\phi_\Omega^{3}~
				D\phi_{\Omega_c}^{1} D\phi_{\Omega_c}^{2} D\phi_{\Omega_c}^{3}~
				\\
				&
								\chi_1^* [\phi_\Omega^1,\phi_{\Omega_c}^1]~
								\chi_2^* [\phi_\Omega^2,\phi_{\Omega_c}^2]~
								\chi_3^* [\phi_\Omega^3,\phi_{\Omega_c}^3]
				\psi_1[\phi_\Omega^2,\phi_{\Omega_c}^1]~
				\psi_2[\phi_\Omega^3,\phi_{\Omega_c}^2]~
				\psi_3[\phi_\Omega^1,\phi_{\Omega_c}^3]~
		\Big]^*
		\\
		&=
		\Big[ 
		\int
		D\phi_{\Omega_c}^{1} D\phi_{\Omega_c}^{2} D\phi_{\Omega_c}^{3}~
		  D\phi_\Omega^{1} D\phi_\Omega^{2} D\phi_\Omega^{3}~
		\\
		&
				\chi_2^* [\phi_{\Omega_c}^2,\phi_\Omega^1]~
				\chi_3^* [\phi_{\Omega_c}^3,\phi_\Omega^2]~
				\chi_1^* [\phi_{\Omega_c}^1,\phi_\Omega^3]~
				\psi_1[\phi_{\Omega_c}^1,\phi_\Omega^1]~
				\psi_2[\phi_{\Omega_c}^2,\phi_\Omega^2]~
				\psi_3[\phi_{\Omega_c}^3,\phi_\Omega^3]~
		\Big]^*
		\\
		&
\text{(redefining $\phi_\Omega^i$ as $\phi_{\Omega_c}^{i-1}$ and thinking of $\phi_{\Omega_c}^i$ as the region of interest.)}		
\\
&=
\langle \chi_2 \chi_3 \chi_1 | E_{\Omega_c}^{(3)} |\psi_1 \psi_2 \psi_3 \rangle^* \, .
	\end{align*}
	In writing $\psi[\phi_\Omega,\phi_{\Omega_c}]$ as $\psi[\phi_{\Omega_c},\phi_\Omega]$ we are just thinking $\mathcal{H}_\Omega \otimes \mathcal{H}_{\Omega_c}$ as $\mathcal{H}_{\Omega_c} \otimes \mathcal{H}_\Omega$, meaning now $\Omega_c$ is our region of interest rather than $\Omega$.
	
	Similar derivation follows for general $n$ and we have
	\begin{align}
	\langle \psi_1 \dots \psi_n | E_\Omega^{(n)} | \chi_1 \dots \chi_n \rangle
	&=
	\langle \chi_{2} \dots \chi_1 | E_{\Omega_c}^{(n)} | \psi_1 \dots \psi_n \rangle^* \, .
	\end{align}

\section{Correction terms} \label{a2}
Now we discuss the leftover ``other pieces" in \ref{correctionterms}. 
	\begin{align}
	\langle 0|
		\tilde{O}_j^\dagger O_i^\dagger E_\Omega^{(n)} O_p \tilde{O}_q |0\rangle
		&=
	\langle 0|
O_i^\dagger E_\Omega^{(n)} O_p 	\tilde{O}_j^\dagger \tilde{O}_q |0\rangle		
+
	\langle 0|
\left[ \tilde{O}_j^\dagger , O_i^\dagger E_\Omega^{(n)} O_p\right] \tilde{O}_q |0\rangle \, .
	\label{even1}
	\end{align}
	$1^{st}$ piece in \ref{even1} can be further simplified as
	\begin{align}
			\langle 0|
			O_i^\dagger E_\Omega^{(n)} O_p 	\tilde{O}_j^\dagger \tilde{O}_q |0\rangle		
			&=
	\langle 0|
	O_i^\dagger E_\Omega^{(n)} O_p 	
	|\alpha\rangle \langle \alpha|
	\tilde{O}_j^\dagger \tilde{O}_q |0\rangle	\, ,		
			\label{even2}
	\end{align}
	where we have inserted a complete set\footnote{For interacting gapless theories there could be confusion regarding which of $\{|\alpha \rangle\}$ represent one particle states. But we are excluding such theories from our analysis.} of states $|\alpha \rangle \langle \alpha |$. Only the $|\alpha\rangle = |0\rangle$ term survives, since for all others terms we have $\langle \alpha|
	\tilde{O}_j^\dagger \tilde{O}_q |0\rangle = 0$. 

	We concentrate on the piece $\langle 0 |
	O_i^\dagger E_\Omega^{(n)} O_p 
	|0\rangle $.
	\begin{align*}
			\langle 0 |
			O_i^\dagger E_\Omega^{(n)} O_p 
			|0\rangle
			&=
			\langle 0 |
			\Big\{ (O_{i_n}^{(n)})^\dagger \dots (O_{i_1}^{(1)})^\dagger\Big\}  E_\Omega^{(n)}
			\Big\{O_{p_1}^{(1)} \dots O_{p_n}^{(n)} \Big\} |0\rangle
			\\
			&=
			\langle \psi_{i_1}, \dots, \psi_{i_n}| E_\Omega^{(n)}
			|\psi_{p_1},\dots,\psi_{p_n}\rangle
			\\
			&=
			\langle \psi_{p_2}, \dots, \psi_{p_1}| E_{\Omega_c}^{(n)}
			|\psi_{i_1},\dots,\psi_{i_n}\rangle^*
			\\
			&=
			\langle 0 |
			\Big\{ (O_{p_1}^{(n)})^\dagger \dots (O_{p_2}^{(1)})^\dagger\Big\}  E_{\Omega_c}^{(n)}
			\Big\{O_{i_1}^{(1)} \dots O_{i_n}^{(n)} \Big\} |0\rangle^*
			\\
			&=
\langle 0 |
\Big\{ (O_{p_1}^{(n)})^\dagger \dots (O_{p_2}^{(1)})^\dagger\Big\}  
\Big\{O_{i_1}^{(1)} \dots O_{i_n}^{(n)} \Big\} |n\rangle^* \langle n| E_{\Omega_c}^{(n)} |0\rangle^*		
\\
&+
			\langle 0 |
			\Big\{ (O_{p_1}^{(n)})^\dagger \dots (O_{p_2}^{(1)})^\dagger\Big\}  
			\Big[ E_{\Omega_c}^{(n)} , 
			\Big\{O_{i_1}^{(1)} \dots O_{i_n}^{(n)} \Big\} \Big] |0\rangle^*				
			\\
			&=
			\langle 0| E_\Omega^{(n)} |0\rangle 
			\delta_{p_2,i_1} \dots \delta_{p_1,i_n} 
			+
			\langle 0 |
			\Big\{ (O_{p_1}^{(n)})^\dagger \dots (O_{p_2}^{(1)})^\dagger\Big\}  
			\Big[ E_{\Omega_c}^{(n)} , 
			\Big\{O_{i_1}^{(1)} \dots O_{i_n}^{(n)} \Big\} \Big] |0\rangle^*	\, .			
	\end{align*}
	Gathering all these, we see
	\begin{align}
	\nonumber
			 \langle 0 |
			 \tilde{O}_j^\dagger O_i^\dagger 
			 E_\Omega^{(n)}
			 O_p \tilde{O}_q |0\rangle 
			 &=
			 	\langle 0| E_\Omega^{(n)} |0\rangle \times
			 	\delta_{p_2,i_1} \dots \delta_{p_1,i_n} 
			 	\times \delta_{j_1 q_1} \dots \delta_{j_n q_n}		
			 	\\
			 	\nonumber
			 	&+
				\langle 0 |
				\Big\{ (O_{p_1}^{(n)})^\dagger \dots (O_{p_2}^{(1)})^\dagger\Big\}  
				\Big[ E_{\Omega_c}^{(n)} , O_i \Big] |0\rangle^*	
				\times
				\delta_{j_1 q_1} \dots \delta_{j_n q_n}	
				\\
				&+
					\langle 0|
					\left[ \tilde{O}_j^\dagger , O_i^\dagger E_\Omega^{(n)} O_p\right] \tilde{O}_q |0\rangle \, .
		\label{allterms}						 		 	
	\end{align}
In right hand side of \ref{allterms}, apart from the first piece, all other pieces contain commutators between fields that are mostly supported in disjoint regions of space and hence are small. Thus the first term is the leading piece. 
	
\section{Fermionic fields} \label{a3}
Now we have to keep track of some signs and have to arrange stuff in terms of anti-commutators. 
We treat the cases of even and odd $n$ separately.

\paragraph{Even n :}
A generic term is still given by \ref{allterms}, so all conclusions remain the same, in particular \ref{proofconjecture}. Only change is that end of the day we have to break up various commutators in terms of anti-commutators. This can easily be done using the
	\begin{align}
	[A, B_1 \dots B_n]
	&=
	\sum_{p=1}^n
	(-1)^{p-1} B_1 \dots B_{p-1}\{A,B_p\} B_{p+1} \dots B_n\, .
	\label{identity1}
	\end{align}

\paragraph{Odd n:}
In this case we use anti-commutators everywhere, because then we can use the folowing identity to break everything up into anti-commutators
	\begin{align}
	\{A, B_1 \dots B_n\}
	&=
	\sum_{p=1}^n
	(-1)^{p-1} B_1 \dots B_{p-1}\{A,B_p\} B_{p+1} \dots B_n\, .
	\label{identity2}
	\end{align}
We break up a generic term as
\begin{align}
	 \langle 0 | \tilde{O}_j^\dagger O_i^\dagger E^{(n)}_\Omega O_p \tilde{O}_q |0\rangle
	&=
- \langle 0 |O_i^\dagger E^{(n)}_\Omega O_p  \tilde{O}_j^\dagger \tilde{O}_q |0\rangle
+
 \langle 0 | \left\{
\tilde{O}_j^\dagger , O_i^\dagger E^{(n)}_\Omega O_p
 \right\} \tilde{O}_q |0\rangle \, .
 \label{f1}
\end{align}
In the first piece of \ref{f1} we insert complete basis and end up with
$\langle 0 |O_i^\dagger E^{(n)}_\Omega O_p  |0\rangle \langle 0| \tilde{O}_j^\dagger \tilde{O}_q |0\rangle$. In this,
\begin{align}
\nonumber
\langle 0 |O_i^\dagger E^{(n)}_\Omega O_p  |0\rangle
&=
\langle 0 | \left( O^{(n)}_{p_1}\right)^\dagger \dots  \left( O^{(1)}_{p_2}\right)^\dagger~ E^{(n)}_{\Omega_c} ~
O^{(1)}_{i_1} \dots O^{(n)}_{i_n} |0\rangle^*
\\
\nonumber
&=
-\langle 0 | \left( O^{(n)}_{p_1}\right)^\dagger \dots  \left( O^{(1)}_{p_2}\right)^\dagger ~
O^{(1)}_{i_1} \dots O^{(n)}_{i_n} ~ E^{(n)}_{\Omega_c} |0\rangle^*
\\
\nonumber
&+
\langle 0 | O^{(n)}_{p_1} \dots O^{(1)}_{p_2} ~
\left\{ E^{(n)}_{\Omega_c} ,~ O^{(1)}_{i_1} \dots O^{(n)}_{i_n}  \right\} |0\rangle
\\
&=
-\langle 0 | E^{(n)}_\Omega |0\rangle
~\delta_{i_1 p_2} \dots \delta_{i_n p_1}
+
\langle 0 | O^{(n)}_{p_1} \dots O^{(1)}_{p_2} ~
\left\{ E^{(n)}_{\Omega_c} ,~ O^{(1)}_{i_1} \dots O^{(n)}_{i_n}  \right\} |0\rangle \, .
\end{align}
Gathering all these, we see
\begin{align}
\nonumber
	\langle 0 | \tilde{O}_j^\dagger O_i^\dagger E^{(n)}_\Omega O_p \tilde{O}_q |0\rangle
	&=
	\langle 0 | E^{(n)}_\Omega |0\rangle
	~\delta_{i_1 p_2} \dots \delta_{i_n p_1}
	\times
	 \delta_{j_1 q_1} \dots \delta_{j_n q_n}
	\\
	\nonumber
	&-
	\langle 0 | O^{(n)}_{p_1} \dots O^{(1)}_{p_2} ~
	\left\{ E^{(n)}_{\Omega_c} ,~ O^{(1)}_{i_1} \dots O^{(n)}_{i_n}  \right\} |0\rangle
	\delta_{j_1 q_1} \dots \delta_{j_n q_n}
	\\
	&
+
\langle 0 | \left\{
\tilde{O}_j^\dagger , O_i^\dagger E^{(n)}_\Omega O_p
\right\} \tilde{O}_q |0\rangle
\end{align}
Since the leading piece remains the same, \ref{proofconjecture} goes through. The details of the corrections terms is different from those in \ref{a2}, although the difference is only quantitative.
\pagebreak


\begin{thebibliography}{99}

\bibitem{EPR}
A. Einstein, B. Podolsky , N. Rosen,
Phys. Rev. {\bf 47}, 777 (1935) .

\bibitem{Ryu:2006bv} 
  S.~Ryu and T.~Takayanagi,
  Phys.\ Rev.\ Lett.\  {\bf 96}, 181602 (2006)
  [hep-th/0603001].
  
\bibitem{Ryu:2006ef} 
  S.~Ryu and T.~Takayanagi,
  JHEP {\bf 0608}, 045 (2006)
  [hep-th/0605073].

\bibitem{Casini:2004bw} 
  H.~Casini and M.~Huerta,
  Phys.\ Lett.\ B {\bf 600}, 142 (2004)
  [hep-th/0405111].
  
\bibitem{casini2}
 H.~Casini and M.~Huerta, 
 J.\ Phys.\ A {\bf 40} (2007) 7031-7036
[cond-mat/0610375]
  
\bibitem{Vidal:2002rm} 
  G.~Vidal, J.~I.~Latorre, E.~Rico and A.~Kitaev,
  Phys.\ Rev.\ Lett.\  {\bf 90}, 227902 (2003)
  [quant-ph/0211074].
  
\bibitem{Kitaev:2005dm} 
  A.~Kitaev and J.~Preskill,
  Phys.\ Rev.\ Lett.\  {\bf 96}, 110404 (2006)
  [hep-th/0510092].
  
\bibitem{Wen}
M.~Levin and X-G.~Wen,
Phys.\ Rev.\ Lett.\ {\bf 96}, 110405 (2006)
 
\bibitem{Caputa:2014vaa} 
  P.~Caputa, M.~Nozaki and T.~Takayanagi,
  PTEP {\bf 2014}, 093B06 (2014)
  [arXiv:1405.5946 [hep-th]].

  
\bibitem{Nozaki:2014hna} 
  M.~Nozaki, T.~Numasawa and T.~Takayanagi,
  Phys.\ Rev.\ Lett.\  {\bf 112}, 111602 (2014)
  [arXiv:1401.0539 [hep-th]].
 
\bibitem{Shiba:2014uia} 
  N.~Shiba,
  JHEP {\bf 1412}, 152 (2014)
  [arXiv:1408.0637 [hep-th]].
  
\bibitem{Hawking}
  S. W. Hawking, Commun. Math. Phys. {\bf 43}, 199 (1975)
  [Erratum-ibid. {\bf 46}, 206 (1976)];
S. W. Hawking, Phys. Rev. D {\bf 14}, 2460 (1976)
  
\bibitem{Mathur:2009hf} 
  S.~D.~Mathur,
  Class.\ Quant.\ Grav.\  {\bf 26}, 224001 (2009)
  [arXiv:0909.1038 [hep-th]].
  
\bibitem{Almheiri:2012rt} 
  A.~Almheiri, D.~Marolf, J.~Polchinski and J.~Sully,
  JHEP {\bf 1302}, 062 (2013)
  [arXiv:1207.3123 [hep-th]].
  
\bibitem{zarnardi}
Paolo Zanardi,
Phys.\ Rev.\ A \textbf{65}, 042101 (2002)

\bibitem{zarnardiandwang}
Paolo Zanardi, Xiaoguang Wang ,
 J.\ Phys.\ A \textbf{35}, 7947 (2002).



\bibitem{eckertetal}
K. Eckert, J. Schliemann, D. Bru{\ss},
M. Lewenstein,
Ann.\ Phys.\ (N.Y.) \textbf{299}, 88 (2002).

\bibitem{yushi}
Yu Shi,
Phys.\ Rev.\ A \textbf{67}, 024301 (2003).

\bibitem{korbiczandlewenstein}
J. K. Korbicz, M. Lewenstein ,
Phys.\ Rev.\ A \textbf{74}, 022318 (2006).

\bibitem{Balachandran:2013hga} 
  A.~P.~Balachandran, T.~R.~Govindarajan, A.~R.~de Queiroz and A.~F.~Reyes-Lega,
  Phys.\ Rev.\ Lett.\  {\bf 110}, no. 8, 080503 (2013)
  [arXiv:1303.0688 [hep-th], arXiv:1205.2882 [quant-ph]].

\bibitem{review1}
Giancarlo Ghirardi, Luca Marinatto, Tullio Weber ,
 J.\ Stat.\ Phys. \textbf{108}, 49 (2002)
[arXiv:0109017v2[quant-ph]].

\bibitem{review2}
Ryszard Horodecki , Pawel Horodecki , Michal Horodecki,
Karol Horodecki ,
Rev.\ Mod.\ Phys. \textbf{81}, 865 (2009)

\bibitem{bennett}
Charles H. Bennett , Herbert J. Bernstein , Sandu Popescu ,
Benjamin Schumacher ,
Phys.\ Rev.\ A \textbf{53}, 2046 (1996).

\bibitem{reeh-schlieder}
H. Reeh , S. Schlieder,
Nuovo Cimento (10), 22:1051-1068, 1961.

\bibitem{haldane}
Hui Li , F. D. M. Haldane, 
Phys.\ Rev.\ Lett. \textbf{101}, 010504 (2008).

\bibitem{grittings-fisher}
J. R. Grittings , A.J. Fisher ,
Phys.\ Rev.\ A. \textbf{66}, 032305 (2001).

\bibitem{wiseman-vaccaro}
H. M. Wiseman , John A. Vaccaro ,
Phys.\ Rev.\ Lett. \textbf{91}, 097902 (2003).

\bibitem{dowling-doherty-wiseman}
M. R. Dowling , A. C. Doherty , H. M. Wiseman,
Phys.\ Rev.\ A. \textbf{73}, 052323 (2006).

\bibitem{lee-kim}
H. W. Lee , J. Kim, 
Phys.\ Rev.\ A \textbf{63}, 012305 (2000).



%
%
%
%
%
%
%
%





%
%
%

%
\end{thebibliography}
\end{document}